\documentstyle[aps,epsf]{revtex} 

\newcommand{\beq}{\begin{equation}}
\newcommand{\eeq}{\end{equation}}
\newcommand{\beqa}{\begin{eqnarray}}
\newcommand{\eeqa}{\end{eqnarray}}
\newcommand{\asl}{$a_{SL}$}
\newcommand{\apks}{$a_{\psi K_S}$}
\newcommand{\app}{$a_{\pi\pi}$}

\newcommand{\afks}{$a_{\phi K_S}$}
\newcommand{\bbar}{$B^0-\bar B^0$}
\newcommand{\goff}{$\Gamma_{12}$}
\newcommand{\moff}{$M_{12}$}
\newcommand{\bsg}{\mbox{$b \to s \gamma$}}

\def\sm{Standard Model}
\def\np{new physics}
\def\BR{{\rm BR}}
\def\np{new physics}

\newcommand{\mot}{$M_{12}$}
\newcommand{\msmot}{$M^0_{12}$}

\newcommand{\betam}{$\tilde\beta$}

\newcommand{\fb}{$\sqrt{B_B}f_B$}

\def\npb#1{Nucl.\ Phys.\ {\bf B #1}}
\def\plb#1{Phys.\ Lett.\ {\bf B #1}}
\def\prd#1{Phys.\ Rev.\ {\bf D #1}}
\def\prl#1{Phys.\ Rev.\ Lett. {\bf #1}}

\def\rmp{Rev.\ Mod.\ Phys.\ }

\def\progtp#1{Prog.\ Th.\ Phys.\ {\bf #1}}
\def\hep#1{hep-ph\ {#1}}
\def\epj#1{Eur.\ Phys.\ J.\ {\bf #1}}

\begin{document}     

\baselineskip 14pt
\title{New Physics in CP Violating $B$ Decays
\footnote{Invited talk at the American Physical Soceity, Division of
Particles and Fields Conference, Jan 5-9 1999, Los Angeles CA.}}
\author{Mihir P. Worah}
\address{
Department of Physics \\
University of California, Berkeley, California 94720 \\
and \\
Theoretical Physics Group \\
Lawrence Berkeley National Laboratory, Berkeley, California 94720 
}

\maketitle             

\vspace{-135pt}
\begin{center}
  \hfill  UCB-PTH-99/19 \\
~{} \hfill    LBNL-43170 \\
~{}  \hfill    May 1999 \\
\end{center}
\vspace{100pt}

\begin{abstract}       

We discuss the sensitivity of the CP violating measurements at the 
upcoming $B$ factories to the presence of physics beyond the \sm. We
review the three manifestations of CP violation possible in the $B$
meson system. We give examples of decay modes for each of these which
are sensitive to new physics, are experimentally feasible, 
and theoretically clean. Finally, we present techniques to
extract \sm\ parameters in the presence of new physics.
\
\end{abstract}   

\section{Introduction}

CP violation has so far only been observed in the decays of neutral
$K$ mesons. It is one of the goals of the proposed $B$ factories
to find and study $CP$ violation in the decays of $B$
mesons, and thus elucidate the mechanisms by which CP
violation manifests itself in the low energy world. 
There is a commonly accepted \sm\ of CP violation,
namely that it is a result of the one physical phase in the $3 \times 3$
Cabbibo Kobayashi Maskawa (CKM) matrix \cite{CKM}. 
If, however, there is physics beyond the \sm, we would expect to see
its effects in CP violating $B$ decays.

An important task when trying to detect new physics is to identify decay
modes where one could find large deviations from the \sm\
expectations. Thus, one needs to find processes that are not
only sensitive to \np, but also  
experimentally accessible and
for which there exist well defined \sm\ expectations.
Moreover, if the presence of new physics is detected, it is then
important to try and disentangle the new physics contributions to the
CP violation from the \sm\ contribution.

CP violation can manifest itself in $B$ decays due to three distinct
mechanisms. ``Indirect CP violation'' which is caused by a phase in
the \bbar\ mixing amplitude. ``Direct CP violation'' which is caused
by interfering decay amplitudes. And finally, ``mixed CP violation'' is
caused due to interference between the \bbar\ mixing amplitude and the 
$B$ decay amplitudes. 
In this talk we give examples of CP violating $B$ decays
for each of these three possible manifestations of CP violation, and
which could allow an early detection of new physics.
These examples were chosen
because they are both experimentally and theoretically ``clean''. 
Finally, we discuss a technique based on measuring the CP violation in 
semi-leptonic $B$ decays that could help separate the \sm\
parameters from the \np\ ones.

\section{CP Violation in $B$ Decays.}

In this section we review the three sources of CP violation in $B$
decays, and give examples for each of these where \np\ could affect
the \sm\ predictions in an observable way.

\subsection{Indirect CP Violation}

This arises due to a phase between $\Gamma_{12}$ and $M_{12}$, the
absorbtive and dispersive parts of the \bbar\ mixing amplitude
respectively. 
It measures the asymmetry in the process 
\beq
B^0 \to \bar B^0 ~~~~~~~~~{\rm vs.}~~~~~~~~~ \bar B^0 \to B^0
\eeq
and is experimentally measured as 
\beq
a_{SL} \equiv \frac{
                              \Gamma(\bar B^0\to l^{+} X) - 
                              \Gamma(B^0 \to l^{-} X)
                             }
                             {
                              \Gamma(\bar B^0\to l^{+} X) + 
                              \Gamma(B^0 \to l^{-} X)
                             },
\eeq
the CP violation in inclusive semi-leptonic $B$ decays. 
The \sm\ expectation for 
\beq
a_{SL} = {\rm Im}(\frac{\Gamma_{12}}{M_{12}})
       = \left |\frac{\Gamma_{12}}{M_{12}} \right |\sin\phi_{12},
\label{a_sl_def}
\eeq
(where $\phi_{12}$ is the phase between $\Gamma_{12}$ and $M_{12}$) 
calculated using local quark-hadron duality, is 
$a_{SL}^{SM} < 10^{-3}$ \cite{Hagelin} which is unobservably
small. Thus, an observation of CP violation in this mode would signal
the presence of physics beyond the \sm.
The smallness of the \sm\ expectation is due to the fact that 
$|\Gamma_{12}/M_{12}| \sim 10^{-2}$ and because the GIM mechanism
results in $\sin\phi_{12} \sim m_c^2/m_b^2 \sim 10^{-1}$.
Thus, new physics can enhance $a_{SL}$ by increasing
$|\Gamma_{12}/M_{12}|$ and/or $\sin\phi_{12}$.

Most models of \np\ introduce new heavy particles that contribute
to \moff\ but not \goff. This could lead to enhancements of
$\sin\phi_{12}$, thus allowing $a_{SL} \sim 0.01$ \cite{Lisa}, 
which would be observable in about one year of running at the $B$
factories.
In order for new physics to significantly affect \goff, one would 
need either large new
decay amplitudes into known states that are common to both $B^0$
and $\bar B^0$, or to introduce additional, exotic common final
states. Such a scenario 
could enhance both the factors mentioned above, and could lead to
$a_{SL} \sim 0.1$ \cite{GPW}. 
This would be detected in the very early stages of
data taking at the asymmetric $B$ factories, with only 
about $10^6$ \bbar\ pairs. 

\subsection{Direct CP Violation}

This form of CP violation arises from the interference between two or more
decay amplitudes for the $B$ mesons to decay to a
particular final state. It could arise in the decays of charged as
well as neutral $B$ mesons, and measures the asymmetry between the rates
for 
\beq
B \to f ~~~~~~~~~{\rm vs.}~~~~~~~~~ \bar B \to \bar f
\eeq
where $f$ is some final state, and $\bar f$ is its CP conjugate.

The \sm\ expectations for this kind of CP
asymmetry in exclusive modes is hard to calculate. This is because in
addition to the well defined CP violating phase between the amplitudes
(arising from the CKM matrix) one also needs to compute the CP
conserving strong interaction phases between these amplitudes. These
can be estimated using any of a number of hadron models, but 
the uncertainties are large and essentially incalculable. Thus, one is 
led to consider inclusive modes, where one can use the notion of 
global quark-hadron
duality to produce reliable \sm\ predictions. One such inclusive
asymmetry is 
\beq
a_{\bsg} \equiv \frac{
                              \Gamma(\bar B\to X_s \gamma) - 
                              \Gamma(B \to X_{\bar s} \gamma)
                             }
                             {
                              \Gamma(\bar B\to X_s \gamma) + 
                              \Gamma(B \to X_{\bar s} \gamma)
                             },
\label{def}
\eeq
the CP asymmetry in the $b \to s\gamma$ decay.  

The \sm\ expectation is $a_{b\to s\gamma} < 0.015$ \cite{KN}, which is 
unobservably small for this mode. 
The presence of new physics could significantly enhance this
asymmetry, leading to $a_{b\to s\gamma} \sim 0.1$ \cite{KN} which
should be observable in the first year at the $B$ factories. Detection 
of a CP asymmetry at this level would be a clear signal of \np.
Note, that the observed 
$\BR(b\to s \gamma)=(3.15\pm 0.54)\times 10^{-4}$ \cite{CLEO} is in good 
agreement with the \sm\ expectation 
$\BR(b\to s \gamma)=(3.29\pm 0.33)\times 10^{-4}$ \cite{MG}. Thus, one
has to ensure that the proposed \np\ effects that contribute to the CP 
asymmetry in this mode interfere destructively in their contribution
to the total decay width for it.

\subsection{Mixed CP Violation}

This is caused due to interference between the amplitude for a $B$ to
decay into some final state $f$ with the amplitude for a $B$ to first
oscillate into a $\bar B$ which subsequently decays to the same final
state $f$. It measures the asymmetry in the process
\beq
B^0 \to \bar B^0 \to f~~~~~~~~~{\rm vs.}~~~~~~~~~ 
\bar B^0 \to B^0 \to \bar f
\eeq
The theoretical predictions are particularly clean when the final
state is a CP eigen state, and there is only one decay amplitude to
that state \cite{Jon}. This is the case for
\beq
a_{\psi K_S} \equiv {\Gamma(B^0\to\psi K_S)-
\Gamma(\bar B^0\to\psi K_S)\over
\Gamma(B^0\to\psi K_S)+
\Gamma(\bar B^0\to\psi K_S)},
\label{bpks}
\eeq
the CP asymmetry in $B \to \psi K_S$
which measures $\sin 2\beta$ in the \sm. 
Interestingly, within the \sm,
\beq
a_{\phi K_S} \equiv {\Gamma(B^0\to\phi K_S)-
\Gamma(\bar B^0\to\phi K_S)\over
\Gamma(B^0\to\phi K_S)+
\Gamma(\bar B^0\to\phi K_S)},
\label{bfks}
\eeq
the CP asymmetry in $B \to \phi K_S$ 
also measures $\sin 2\beta$ to a high degree of accuracy \cite{LP,GIW}. 
Since $B_d \to \phi K_S$ is a loop mediated 
process within the \sm, it is not unlikely that new physics could
have a significant effect on it \cite{GW}. 
The expected branching ratio and the high identification
efficiency for this decay suggests 
that \afks\ is experimentally accessible at the early stages of 
the asymmetric $B$ factories. 
Thus, the search for a difference between \apks\ and \afks\ is a 
promising way to look for physics beyond the \sm\ 
\cite{GW,Others}. A difference $|a_{\psi K_S}-a_{\phi K_S}| > 5\%$
would be an indication of \np. A similar analysis can be carried out
for $a_{\eta' K_S}$, the CP asymmetry in $B \to \eta' K_S$ \cite{LS}.

\section{Separating the new physics from the Standard Model}

Most models of physics beyond the \sm\ only affect the \bbar\ mixing
amplitude $M_{12}$ without significantly affecting the $B$ decay
amplitudes. 
In that case, one can couple the already measured 
values of $|V_{ub}|$ and $\Delta m_B$ 
with the measurements of \apks\ and \app, the CP violating asymmetries 
in the decays $B\to \psi K_S$ and $B \to \pi\pi$ respectively, to 
disentangle the new physics contributions to \bbar\ mixing from
the \sm\ ones \cite{GNW}. 
A shortcoming of this approach is that discrete ambiguities in
relating \apks\ and \app\ to CKM phases leads to multiple solutions
for the \sm\ and new physics parameters \cite{GNW,GQ}.
Thus, one needs additional
information to try to resolve these.

Here we use a graphical representation of the data in the \moff\ plane 
\cite{Goto} to highlight the information that can
be obtained from a measurement of \asl, the CP violation in semi-leptonic
$B$ decays. 
The sensitivity of \asl\ to new physics has already been discussed in
the previous section.
We show, in addition, how one can use constraints on, or the observation
of, \asl\ to restrict allowed regions in the \sm\ parameter space \cite{CW}.

\subsection{The complex $M_{12}$ plane}

Under the assumption that the $B$ decay amplitudes are not affected,
all the new physics effects can be expressed in terms of one complex
number: the new contribution to the dispersive part of the 
\bbar\ mixing amplitude, $M_{12}$. Explicitly, we write 
\beq
M_{12} = r^2 e^{i2\theta}M_{12}^0
\label{m12_def}
\eeq
where $M_{12}^0$ represents the Standard Model contribution. 
We will work in the convention where the phase of $M_{12}^0$ 
is $2\beta$, thus that of \moff\ is 
$2(\beta+\theta) \equiv 2\tilde\beta$. 
(Note, that these phases are 
measured relative to that of the $b \to c \bar c d$ decay amplitude).

The magnitude of $M_{12}$ is well determined:
\beq
|M_{12}|=\Delta m_B/2
\eeq
where  $\Delta m_B= 0.470\pm 0.019$ ps$^{-1}$ =$3.09\times 10^{-13}$
 GeV\ \cite{PDG}. 
We can use this to represent the actual value of \mot\ 
as lying somewhere on the unit circle centered at the origin of the
complex \moff\ plane (where
all data are rescaled by the experimentally determined central value of
$\Delta m_B/2$). The phase of \moff, $2\tilde\beta$, 
will be obtained from the CP asymmetry in $B \to \psi K_S$:
\beq
a_{\psi K_S} = \sin 2\tilde\beta.
\eeq

\begin{figure}[ht]
\centerline{\epsfxsize 3.0 truein \epsfbox{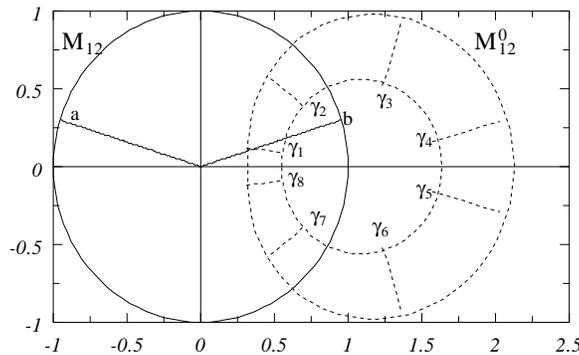}}   
\caption[]{
\label{gamma_8}
\small The complex $M_{12}$ plane in units of $\Delta m_B/2$.
We show the two-fold discrete ambiguity in the value of $2\tilde\beta$
(the points $a$ and $b$)
and the eight-fold ambiguity in $\gamma$ 
(the curves labeled $\gamma_1 ... \gamma_8$)
resulting from the
measurements \apks = 0.3 and \app = -0.7.
We have used $\sqrt{B_B}f_B= 200$ MeV in obtaining the Standard Model
region.}
\end{figure}

We can plot the allowed \sm\ region in this plane using \cite{BBH}
\beq
M_{12}^0 = \frac{\Delta m_B}{2}\left | \frac{V_{tb}V_{td}^*}{0.0086}
\right |^2 \left ( \frac{\sqrt{B_B}f_B}{200~{\rm MeV}} \right )^2
e^{2i\beta}
\label{m120_1}
\eeq
In the absence of new physics, $M_{12}^0=M_{12}$ and one can directly
use $\Delta m_B$ to infer a value for $|V_{tb}V_{td}^*|$.  Although
this is not possible if new physics is present, we can still use the 
unitarity of the CKM matrix to plot an allowed region for the 
\sm, and thus constrain $|V_{tb}V_{td}^*|$. Using 
${V_{ub}/V_{cb}}=ae^{-i\gamma}$ with 
$0.06 \leq a\leq 0.10$
and considering $V_{ud}= 0.975$, $V_{cd} = -0.220$, and 
$V_{cb}=0.0395$ \cite{PDG}
as well determined relative to the other uncertainties in the
problem, we obtain
$
|V_{tb}V_{td}^*|e^{-i\beta} = -0.0395(-0.220 + 0.975ae^{-i\gamma}).
$
Using this relation in Eq. (\ref{m120_1}), we find that
as $a$ covers the stated range and $\gamma$ varies over $0$
to $2\pi$, \msmot\ covers a region of the complex $M_{12}$ plane
as shown in Fig.~\ref{gamma_8}.
\moff, the full \bbar\ mixing amplitude can lie anywhere on the solid
circle, and $M_{12}^0$, the \sm\ contribution lies somewhere in the
region between the two dashed curves. If there were no new physics,
\moff\ would have to lie on the solid circle in one of the two regions
where it intersects with the allowed \sm\ area.

Measuring \app, the CP asymmetry in $B\to \pi\pi$, would give
$\sin 2(\gamma+\tilde\beta)$ (once the penguin effects are determined)
Since, in principle, both \betam\ and $\gamma+\tilde\beta$ are known,
$\gamma$ itself is known. 
Thus, in principle, the CP violating measurements \apks\ and \app\
allow us to disentangle the \sm\ contribution to \bbar\ mixing from the 
new physics contribution. 

Without additional inputs, however, the measurements of \apks\ and \app\ only
allow us to extract $2\tilde\beta$ up to a two-fold ambiguity, and $\gamma$
up to an eight-fold ambiguity as shown in Fig. \ref{gamma_8}.
The true value of the
\bbar\ mixing amplitude, $M_{12}$ could be either of the points
labeled $a$ or $b$. The \sm\ contribution to it, $M_{12}^0$ could lie
on any one of the curves labeled $\gamma_1$ through $\gamma_8$. 
Although there exist techniques that allow a direct extraction of the
angle $\gamma$, these are either experimentally difficult \cite{GLW},
or suffer from theoretical uncertainties and sensitivity to new
physics \cite{FM}. We will now
discuss how a measurement of, or constraints on \asl\ 
restricts the allowed \sm\ parameter space and helps 
resolve some of these discrete ambiguities.

\subsection{The role of \asl}

\begin{figure}[ht]
\centerline{\epsfxsize 3.0 truein \epsfbox{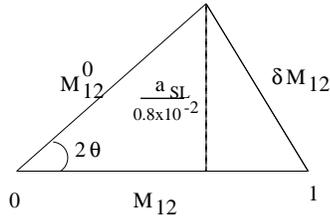}}   
\caption[]{
\label{triangle}
\small The relationship between $M_{12}$, $M_{12}^0$, and $a_{SL}$.
The perpendicular distance between $M_{12}$ and the tip of the
$M_{12}^0$ vector is given by $a_{SL}/0.8\times 10^{-2}$. Where
$0.8\times 10^{-2}$ is the calculated central value of
$\Gamma_{12}^0/M_{12}^0$.  
}
\end{figure}

Within the \sm, at leading order we have \cite{Hagelin}
\beq
\frac{\Gamma_{12}^0}{M_{12}^0} = -5.0\times 10^{-3}
                                 \left (1.4\frac{B_S}{B_B} + 0.24 +
                                 2.5 \frac{m_c^2}{m_b^2}
                                 \frac{V_{cb}V_{cd}^*}{V_{tb}V_{td}^*}
                                 \right ).
\label{sm_asl}
\eeq
where $B_S$ and $B_B$ are the 
bag factors corresponding to the matrix elements of the operators 
$Q_S \equiv (\bar b d)_{S-P}(\bar b d)_{S-P}$ and
$Q \equiv (\bar b d)_{V-A}(\bar b d)_{V-A}$. 
In the vacuum saturation
approximation one has $B_S/B_B=1$ at some typical hadronic scale, and
this expectation is confirmed by a leading order lattice calculation
\cite{Pierro}.
From the measured value of $|V_{ub}/V_{cb}|$ and 
CKM unitarity we know that $|\sin\beta| < 0.35$.
Then, using $m_c^2/m_b^2=0.085$ 
and ${\rm Im}(V_{cb}V_{cd}^*/V_{tb}V_{td}^*) \sim \sin\beta$ 
leads to the limit ${\rm Im(\Gamma_{12}^0/M_{12}^0)} = 
a_{SL}^{SM} < 10^{-3}$ which is unobservably
small. To simplify matters, we will ignore this small phase in the
\sm\ value of \goff/\moff.
One can then write
\beqa
\frac{\Gamma_{12}}{M_{12}} &=& \frac{\Gamma_{12}}{M_{12}^0}
                              \frac{M_{12}^0}{M_{12}} \nonumber \\ 
                     &=& -0.8 \times 10^{-2}\frac{e^{-i2\theta}}{r^2}
\label{g_by_m}
\eeqa
where we have used Eq. (\ref{m12_def}) in
Eq. (\ref{sm_asl}).
Thus, Eqs. (\ref{a_sl_def}) and (\ref{g_by_m}) lead to
\beqa
a_{SL} &=&  0.8 \times 10^{-2}
          {\rm Im}(\frac{M_{12}^0}{M_{12}}) \nonumber \\ 
      &=& 0.8 \times 10^{-2} \frac{\sin 2\theta}{r^2}
\label{asl}
\eeqa
Combining Eqs. (\ref{m12_def}) and (\ref{asl}) one sees that 
$M_{12}^0$ is given by a vector
at an angle $2\theta$ from $M_{12}$ and whose tip is a perpendicular 
distance $a_{SL}/0.8\times10^{-2}$ from it.
In Fig. ~\ref{triangle} we demonstrate this
relation between \moff, \msmot, and \asl. 

\begin{figure}[ht]	
\centerline{\epsfxsize 3.0 truein \epsfbox{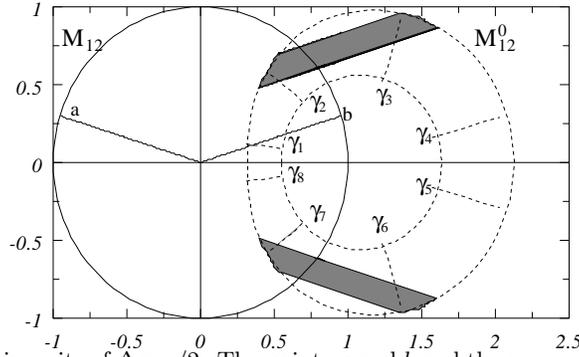}}   
\caption[]{
\label{final2}
\small The complex $M_{12}$ plane in units of $\Delta m_B/2$.
The points $a$ and $b$ and the curves $\gamma_1...\gamma_8$ result
from the measurements \apks = 0.3 and \app = -0.7.
The shaded region corresponds to the allowed Standard Model parameter
space coming from a measurement of $a_{SL}=(-5\pm 1)\times 10^{-3}$.
We have used $\sqrt{B_B}f_B= 200$ MeV in obtaining the Standard Model
region.}
\end{figure}

In Fig. \ref{final2} we 
use a hypothetical scenario to highlight the effects 
of combining \apks\ and \app\ 
with \asl\ in constraining the allowed \sm\ parameter space. 
As before, we use \fb\ = 200 MeV and $0.06 \le a \le 0.10$ 
to construct the allowed \sm\  region, and assume that 
$a_{\psi K_S} = 0.3$, and $a_{\pi\pi}=-0.7$
have been measured. 
We then assume a measurement of $a_{SL} = (-5 \pm 1) \times
10^{-3}$. In this case the \sm\ point must lie in one of the two
shaded bands parallel to the \moff\ vectors $a$ and $b$ respectively. 
For particular values of $\gamma$, this construction gives us 
both $\sin 2\theta$ and $r^2$, hence one has not only resolved 
the \sm\ parameters, but also the new physics ones.
Note, however, that we have ignored the uncertainties in the
calculation of Eq.~(\ref{sm_asl}). These uncertainties, and their 
effects on the analysis presented here are discussed in Ref.~\cite{CW}.

\section{Conclusions}

We have reviewed the three different ways that CP violation can
manifest in the $B$ meson system. We have given example of decay modes 
for each of these classes of CP violation that could allow an early
detection of new physics effects. Finally, we have discussed a
technique based on the CP violating asymmetry in semi-leptonic $B$
decays that helps us in separating the \sm\
contributions from the \np\ ones.

{\em Acknowledgements} The author wishes to thank R. Cahn and
Y. Grossman for collaboration on some of the work presented here. This 
work was supported by the National Science Foundation under grant
PHY-95-147947.


\begin{references} 

\bibitem{CKM}
N. Cabibbo, \prl{10} (1963) 531; 
M. Kobayashi and T. Maskawa, \progtp{49} 652 (1973).

\bibitem{Hagelin}
J. Hagelin, \npb{193}, 123 (1981);
A. Buras, W. Slominski and H. Steger, \npb{245}, 369 (1984);
M. Beneke, G. Buchalla and I. Dunietz, \prd{54}, 4419 (1996).

\bibitem{Lisa}
A. Sanda and Z. Xing, \prd{56}, 6866 (1997);
L. Randall and S. Su, \npb{540}, 37 (1999).

\bibitem{GPW}
Y. Grossman, J. Pelaez and M. Worah, \prd{58}, 096009 (1998).

\bibitem{KN}
A. Kagan and M. Neubert, \prd{58} 094012 (1998).

\bibitem{CLEO}
The CLEO collaboration, preprint CLEO-CONF 98-17, (1998).

\bibitem{MG}
K. G. Chetyrkin, M. Misiak and M. M. Munz, \plb{400} 206 (1997);
C. Greub and T. Hurth \prd{56} 2934 (1997);
A. Kagan and M. Neubert, \epj{7} 5 (1999).

\bibitem{Jon}
I. Dunietz and J. Rosner, \prd{34} 1404 (1986).

\bibitem{LP}
D. London and R. Peccei, \plb{223} 163 (1989).

\bibitem{GIW}
Y. Grossman, G. Isidori and M. Worah, \prd{58} 057504 (1998).

\bibitem{GW}
Y. Grossman and M. Worah, \plb{395} 241 (1997).

\bibitem{Others}
M. Ciuchini {\em et al.}, \prl{79} 978 (1997);
D. London and A. Soni, \plb{407} 61 (1997);
R. Barbieri and A. Strumia, \npb{518} 714 (1998).

\bibitem{LS}
D. London and A. Soni, \plb{407} 61 (1997).

\bibitem{GNW}
Y. Grossman, Y. Nir and M. Worah, \plb{407} 307 (1997).

\bibitem{GQ}
Y. Grossman and H. Quinn, \prd{56} 7259 (1997).

\bibitem{Goto}
T. Goto {\em et al.}, \prd{53} 6662 (1995).

\bibitem{CW}
R. Cahn and M. Worah, LBNL-43159; \hep{9904480}.

\bibitem{PDG}
Review of Particle Properties, \epj{3} 1 {1998}.

\bibitem{BBH}
A. Buras, G. Buchalla and M. Harlander, \rmp{68} 1125 (1996).

\bibitem{GLW}
M. Gronau, and D. London, \plb{253} 483 (1991); 
M. Gronau and D. Wyler, \plb{265} 172 (1991);
D. Atwood, I. Dunietz and A. Soni, \prl{78} 3257 (1997).

\bibitem{FM}
R. Fleischer and T. Mannel, \prd{57} 2752 (1998);
M. Neubert and J. Rosner, \plb{441} 403 (1998).

\bibitem{Pierro}
M. Di Pierro, private communication.

 \end{references}
\end{document}